# Equilibrium microphase separation in the two-leaflet model of lipid membranes

Ramon Reigada[1] and Alexander S. Mikhailov[2,3]

[1]Departament de Química Física i Institut de Química Teòrica i Computacional (IQTCUB), Universitat de Barcelona, Avinguda Diagonal 647, 08028 Barcelona, Spain
e-mail address: reigada@ub.edu

[2]Abteilung Physikalische Chemie, Fritz-Haber-Institut der Max-Planck-Gesellschaft, Faradayweg 4-6, 14195 Berlin, Germany

[3]Department of Mathematics and Life Sciences, Hiroshima University, 1-3-1 Kagamiyama, Higashi-Hiroshima, 739-8530, Japan

A novel two-leaflet description of lipid membranes is proposed. Within its framework, phase separation phenomena in multicomponent biological membranes are analyzed. As we show, interactions between the leaflets tend to suppress macroscopic phase segregation (i.e., complete demixing of lipids) in such systems. Instead, microphase separation characterized by formation of equilibrium nanoscale domains can take place. The phase diagram is constructed and numerical simulations revealing nanostructures of different morphology are performed.

**PACS:** 82.20.Wt, 82.45.Mp, 47.54.-r

Much effort has been devoted to understand how the plasmatic cell membrane organizes its components because this organization determines the functionality of the membrane [1]. Biological lipid bilayers usually include surface domains with different chemical composition. While macroscopic phase separation accompanied by complete demixing of lipids can be indeed observed in simple model bilayers [2], the view of finite-size lipid domains on nanometer scales is broadly accepted in biology. Still controversial, the picture of lipid rafts [3] provides a ground for understanding the relationship between the structure and the function in cellular membranes [4]. Even though substantial experimental data is available, the physical origin of lipid rafts remains under debate [5].

There are two general scenarios that account for prevention of complete phase separation in binary solutions and development of stationary finite-size domains. At equilibrium, such domains may arise from the competition between attractive local and repulsive long-ranged interactions between particles [6]. In nonequilibrium systems, microphase separation can be observed when energetically activated reactions between two components are included [7,8]. Mathematically, the Turing bifurcation [9] underlies both situations.

The nonequilibrium scenario has inspired a variety of proposals in the context of the cell membrane: cholesterol exchange [10,11], action of externally-induced reactions [12], cytoskeletal activity [13, 14] and other active cellular processes [15] have been invoked. However, specific equilibrium mechanisms leading to microphase separation and formation of stable rafts have not so far been proposed.

In this Rapid Communication, we investigate a model of a phase segregating bilayer where microphase separation is possible under equilibrium conditions. In contrast to the classical Helfrich description [16], two leaflets, each with its local curvature and lipid composition, are explicitly considered and interactions tending to minimize the local distance between the leaflets are introduced. Previously, two-leaflet models have been employed to describe fluctuations in membrane thickness [17]. Here, multicomponent two-leaflet membrane systems are explored. Repulsive longed-ranged interactions between lipids are absent in these systems, but macroscopic phase separation can nonetheless be suppressed due to membrane thickness restriction effects.

In our model, the membrane is described as consisting of two coupled laterally heterogeneous elastic surfaces (leaflets) that interact one with another. Close to the critical point, the Landau free energy (in $k_B T$ units) associated with phase segregation can be expressed in terms of composition order parameters $(\phi_+, \phi_-)$ representing local differences in lipid concentrations for each leaflet,

$$\Im_{seg} = \int dr^2 \left[ -\frac{\alpha}{2}\phi_+^2 + \frac{\beta}{4}\phi_+^4 + \frac{\gamma}{2}|\nabla\phi_+|^2 - \frac{\alpha}{2}\phi_-^2 + \frac{\beta}{4}\phi_-^4 + \frac{\gamma}{2}|\nabla\phi_-|^2 \right] \quad (1)$$

For positive parameters $\alpha$ and $\beta$, two phases segregate and different local curvatures develop. Instead of using the Helfrich elastic free energy of the membrane as a single sheet, we consider the two leaflets separately, so that

$$\Im_{elastic} = \frac{1}{2}\int d^2 r \left[ \sigma(\nabla h_+)^2 + \kappa(\nabla^2 h_+ - c_{0,+})^2 + \sigma(\nabla h_-)^2 + \kappa(\nabla^2 h_- - c_{0,-})^2 \right] \quad (2)$$

where $h_+$ and $h_-$ are height variations for the upper and lower layers (Fig. 1), $\sigma$ is the surface tension parameter and $\kappa$ is the layer bending modulus. Spontaneous curvatures of each leaflet ($c_{0,+}, c_{0,-}$) are determined by their local lipid composition, i.e. we have $c_{0,+} = \phi_+ c_0$ and $c_{0,-} = -\phi_- c_0$. This means that since the preferred curvature of a particular lipid phase depends on the shape of the predominant lipid species, it changes its sign from one layer to the other (Fig. 1).

The two layers are coupled by an interfacial surface tension term that accounts for the energy between two different lipid domains at opposite leaflets,

$$\mathfrak{I}_{int} = \frac{1}{2}\int d^2r \left[\xi(\phi_+ - \phi_-)^2\right] \quad (3)$$

Furthermore, there is a harmonic energy penalty for the thickness fluctuations defined as half the deviation from the average membrane thickness [17]

$$\mathfrak{I}_{thick} = \frac{1}{2}\int d^2r \left[\chi\left(\left(\frac{h_+ - h_-}{2}\right) - d_0\right)^2\right] \quad (4)$$

where the parameter $\chi$ characterizes the thickness stiffness of the membrane.

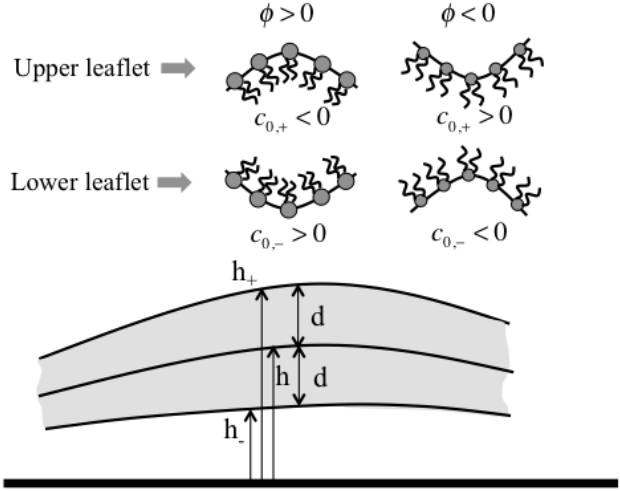

FIG. 1. (Top) Opposite spontaneous curvatures in two membrane leaflets. (Bottom) Local height and thickness variables that describe the shape of a two-leaflet membrane.

For simplicity, only phase segregation processes that take place symmetrically in both leaflets will be considered, so that only one variable $\phi = \phi_+ = \phi_-$ is required and the interfacial free energy (3) vanishes.

Our primary interest in this study is to consider equilibrium thermodynamic structures. Nonetheless, in numerical simulations and in the linear stability analysis, kinetic equations will also be employed. We adopt a conserved kinetic scheme for the composition order parameter,

$$\dot{\phi} = D\nabla^2\left[\frac{\delta\mathfrak{I}}{\delta\phi}\right] \quad (5)$$

Evolution of the membrane shape is described by kinetic equations for the two heights,

$$\dot{h}_+ = -\Lambda\frac{\delta\mathfrak{I}}{\delta h_+} \quad \dot{h}_- = -\Lambda\frac{\delta\mathfrak{I}}{\delta h_-} \quad (6)$$

Relaxation processes of the membrane shape are relatively complex because hydrodynamic solvent flows become induced [18,19]. As result, the relaxation rate constant depends on the wavenumber as $\Lambda(q) = (4\eta q)^{-1}$ where $\eta$ is the kinematic viscosity of the solvent. Here the wavenumber dependence is however neglected since our focus is on final equilibrium patterns that cannot depend on kinetic relaxation details.

The model can be further simplified introducing the midplane height $h$ and the thickness variable $d$ instead of the two leaflet height variables $h_+ = h + d$ and $h_- = h - d$ (Fig.1). The final kinetic equations have the form

$$\dot{\phi} = D\nabla^2\left[\left(2\kappa c_0^2 - \alpha\right)\phi + \beta\phi^3 - \gamma\nabla^2\phi - 2\kappa c_0\nabla^2 d\right]$$
$$\dot{d} = -\Lambda\left[2\kappa\nabla^4 d - 2\sigma\nabla^2 d - \kappa c_0\nabla^2\phi + \chi(d - d_0)\right] \quad (7)$$
$$\dot{h} = -2\Lambda\left[\kappa\nabla^4 h - \sigma\nabla^2 h\right]$$

Note that the midplane height is decoupled from the other two variables and thus the membrane independently tends to become flat.

The linear stability analysis of the stationary solutions $\phi(r) = \overline{\phi}$ and $d(r) = d_0$ with respect to small perturbations $\delta\phi, \delta d \propto \exp[\omega(q)t + iqr]$ can be obtained straightforwardly from the linearized equations and the corresponding linearization matrix

$$L = \begin{pmatrix} q^2\left[\alpha - 2\kappa c_0^2 - \gamma q^2\right] & -2\kappa c_0 q^4 \\ -2\Lambda\kappa c_0 q^2 & -\Lambda\left[\chi + 2\kappa q^4 + 2\sigma q^2\right] \end{pmatrix}$$
(8)

For a tensionless membrane with $\sigma = 0$, a Turing bifurcation is obtained. The first unstable mode has the wavenumber

$$q_0^2 = \frac{2\alpha - 3\kappa c_0^2 \pm c_0\sqrt{\kappa(9\kappa c_0^2 - 4\alpha)}}{2\gamma} \quad (9)$$

and the instability is found at the bifurcation boundary determined by

$$\chi_0^\pm = \frac{\kappa}{\gamma^2}\left[9\kappa c_0^2(2\alpha - 3\kappa c_0^2) - 2\alpha^2 \pm 8\sqrt{\kappa c_0^2\left(\frac{9}{4}\kappa c_0^2 - \alpha\right)^3}\right] \quad (10).$$

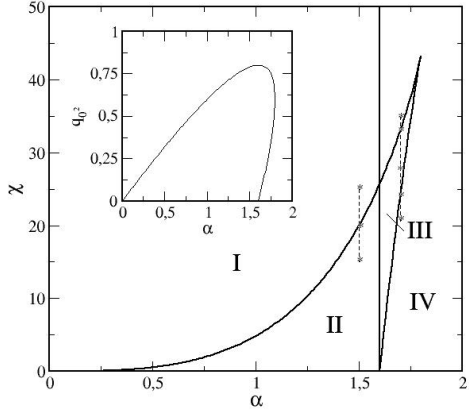

FIG. 2. The phase diagram in the plane $(\chi,\alpha)$ for $\kappa = 20$, $c_0 = 0.2$ and $\gamma = 1$. Region I: no phase separation, region II: microphase separation, region III: coexistence of micro- and macrophase separation, region IV: macroscopic phase separation (spinodal decomposition). (Inset) The wavenumber of the Turing bifurcation critical mode as a function of $\alpha$.

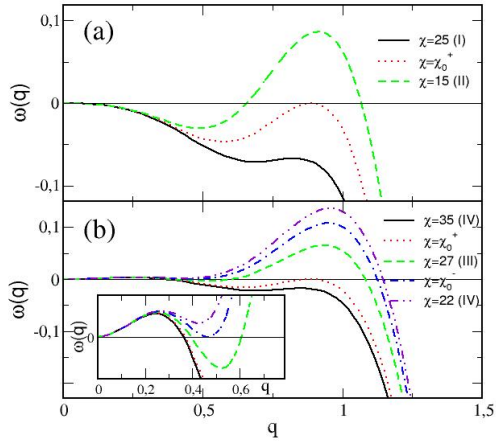

FIG. 3. Increments of growth $\omega$ as functions of the wavenumber $q$ at different values of parameter $\chi$ for (a) $\alpha = 1.5$ and (b) $\alpha = 1.7$. The choices of $\chi$ correspond to the vertical dashed lines in Fig. 2. (Inset) Enlargement of panel (b) for small wavenumbers. Other parameters are $\kappa = 20$, $c_0 = 0.2$ and $\gamma = 1$.

The phase diagram in the $(\chi,\alpha)$ parameter space is displayed in Fig. 2. Characteristic dependences of the increments of growth on the wavenumber $q$ in different parts of the phase diagram are shown in Fig. 3. The cusp separating regions II and III from regions I and IV is formed by the curves $\chi = \chi_0^+(\alpha)$ and $\chi = \chi_0^-(\alpha)$ determined by equation (10). The vertical line, separating regions I and IV and regions II and III, lies at $\alpha = \alpha_{seg} = 2\kappa c_0^2$. Macroscopic phase separation (i.e., spinodal decomposition) takes place in region IV, whereas microphase separation is predicted in region II. The uniform state is stable and phase separation is absent in region I. The linear stability analysis predicts for region III that the spinodal decomposition and microphase separation should coexist (Fig. 3b). Nonlinear numerical simulations in this region reveal (see below) however that spinodal decomposition phenomenon similar to that in region IV finally sets after initial transients.

If thickness stiffness $\chi$ of the bilayer is high, macroscopic phase separation is found above $\alpha = \alpha_{seg}$, whereas the uniform state is stable below this boundary. This limit corresponds to the classical Helfrich description: the thickness of the membrane cannot significantly change and shapes of the two leaflets are almost indistinguishable. When the parameter $\chi$ is decreased, the two leaflets are however less strongly bound one to another and local membrane thickness may substantially vary. According to our results, classical spinodal decomposition should then be preceded by microphase separation where equilibrium domains of fixed sizes are formed (region II).

To determine final equilibrium patterns in different regions of the phase diagram, numerical simulations of equations (7) have been performed. A grid with 100x100 points with periodic boundary conditions was used. The mesh size was $\Delta x = 1$, spatial derivatives were calculated employing a simple centered scheme, and a first order Euler algorithm with time step $\Delta t = 0.0005$ was used for temporal integration. Simulations were started from a uniform state with $\phi(r) = \bar{\phi} = 0, d(r) = d_0$ that was randomly perturbed by applying small local variations.

A uniform stable phase and macroscopic phase separation could be confirmed for regions I and IV. In region II, microphase segregation was getting rapidly achieved. In region III, a pattern of finite-wavelength domains was developing first. Later, however, it was replaced by large homogeneous domains and spinodal decomposition was finally found (Fig. 4a).

Figures 4(b,c) show two typical morphologies (lamellae and spots) of asymptotic patterns observed under microphase separation conditions in numerical simulations for region II. Within each pattern, both local composition and membrane thickness were modulated, as illustrated in Fig. 4d where profiles of these variables in the one-dimensional simulation with the same parameters as in Fig. 4b are displayed. We checked that the characteristic length of these equilibrium patterns was close to the wavelength of the critical Turing mode in Eq. (9), $\sim \pi q_0^{-1}$. Patterns with characteristic length scales in the interval from 5 to 20 were found for $0.15 < \alpha < \alpha_{seg}$.

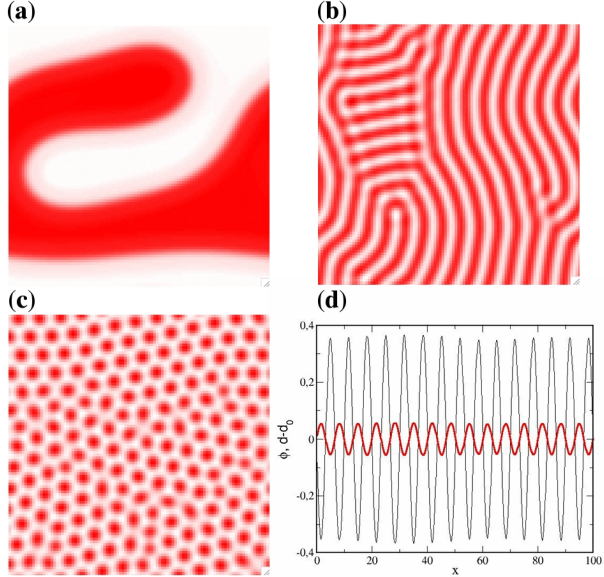

FIG. 4. (a-c) Examples of equilibrium patterns for membrane composition $\phi$: (a) complete phase segregation (region IV, $\alpha = 1.75$, $\chi=35$, $\bar{\phi}=0$), (b) lamellae (region II, $\alpha = 1.5$, $\chi=15$, $\bar{\phi}=0$), (c) spots $\alpha = 1.5$, $\chi=15$, $\bar{\phi}=0.15$). Panel (d) shows profiles of composition order parameter (thick) and thickness variation (thin) in a one-dimensional simulation for $\alpha = 1.5$, $\chi=15$, $\bar{\phi}=0$. Other parameters are the same as in Fig. 1. The linear size of the system is 100 dimensionless length units corresponding to 200 nm.

The correspondence of the model units with real units is provided once the model parameters are fixed to realistic values. For membranes composed of phospholipids, bending rigidities $\kappa$ of few tens of $k_BT$ are typical [20] and for cell membranes containing large molar fractions of cholesterol even higher bending rigidities are possible [21]. Hence, values of $\kappa$ between 10 and 30 are realistic. The parameter $\chi$ could previously be estimated from molecular dynamics simulations as $\chi = 4\times 10^{-21} J/nm^4 \approx 1 k_BT/nm^4$ J/nm$^4$ [17]. According to the Cahn-Hilliard theory [22], $\gamma \approx u l_0^2/2$ where the interaction energy is about $u = 1\, k_BT$ and the characteristic interface width is $l_0 = 2-3$ nm. By choosing $\gamma = 1$, the length unit in our equations becomes fixed to about $l_u = 2$ nm, thus implying $\chi \approx 16\, k_BT/l_u^4$. For the spontaneous curvature radius, estimates between 2 and 20 nm (namely, between 1 and 10 $l_u$) are available [23], so that $c_0 = 0.1-1\, l_u^{-1}$. Requiring that the diffusion constant is $D = 1\, l_u^2/t_u = 1\, \mu m^2/s$, the time unit in our model is fixed to $t_u = 4\times 10^{-6}$ s.

Under these numerical values, region II in the phase diagram (Fig. 2) is realistically accessible and equilibrium microphase separation with characteristic length scales about a few tens of a nanometer can be found. Note that these soft nanostructures are sensitive to thermal fluctuations and, instead of a perfect array, irregular and fluctuating patterns of stripes or spots should actually be expected in such regimes.

The basic mechanism of microphase separation can also be revealed by a simple analysis. Applying equations (1), (2) and (4) to modulated periodic phases with wave numbers q and amplitudes $\phi_m, d_m$ for the composition and thickness variables respectively, the following expression for the free energy is obtained:

$$\Im(q) = -\frac{\alpha}{4}\phi_m^2 + \frac{3\beta}{16}\phi_m^4 + \frac{\gamma}{4}\phi_m^2 q^2 + \frac{\chi}{4}d_m^2 + \kappa\left(d_m q^2 + c_0 \phi_m\right)^2 \quad (11)$$

In the limit of large $\chi$, variations of membrane thickness are suppressed and the minimum of free energy (11) is achieved at $q \to 0$. On the other hand, for small $\chi$, membrane thickness can be modulated and the equilibrium is reached when the energy associated with the curvature [the last term in equation (11)] is minimized, leading to $d_m = -c_0 \phi_m/q^2$. Then, the free energy (11) reads

$$\Im(q) = -\frac{\alpha}{4}\phi_m^2 + \frac{3\beta}{16}\phi_m^4 + \frac{\gamma}{4}\phi_m^2 q^2 + \frac{\chi c_0^2}{4q^4}\phi_m^2 \quad (12)$$

and its minimum is achieved for a modulated phase with the finite wavenumber $q = \left(2\chi c_0^2/\gamma\right)^{1/6}$. Note that this agrees with equations (9) and (10) that yield the same critical wavenumber on the instability boundary when $\alpha$ (and therefore $\chi$) are small. Thus, either macro- or microphase separation take place depending on the competition between domain coarsening and thickness modulation. At short wavelengths, the preferred curvature can be locally achieved without implying large thickness energy penalties.

In summary, we have constructed a simple two-leaflet model for a bilayer membrane with two lipid components and demonstrated that microphase segregation is possible in this model under equilibrium conditions if local membrane thickness can be modulated. For plausible values of bilayer rigidities and curvatures, characteristic sizes of predicted stable structures lie on the nanoscale. These results appear to agree with the experimental observations of lipid rafts on the plasma cell membrane whose existence has been so far attributed to nonequilibrium conditions. Although the nonequilibrium nature of biological membranes indicates that activated processes are important to explain their lateral organization, we have demonstrated that spontaneous formation of equilibrium nanostructures in the cell membrane is also possible, thus suggesting that equilibrium aspects based on membrane thickness modulation have to be taken into account. It would be interesting to verify such predictions in direct MD simulations and in the experiments.

This work has been supported by SEID through project BFU2010-21847-C02-02.